# Conversational Concurrency with Dataspaces and Facets

Sam Caldwell[b] , Tony Garnock-Jones[a] , and Matthias Felleisen[b]

a   Maastricht University, Maastricht, Netherlands
b   Northeastern University, Boston, USA

**Abstract**
**Context**   Developers have come to appreciate the simplicity of message-passing actors for concurrent programming tasks. The actor model of computation is easy to grasp; it is just a conversation among actors with a common goal. Importantly, it eliminates some basic pitfalls of the dominant shared-memory model, most critically data races.
**Inquiry**   A close look at real-world conversations suggests, however, that they are not mere exchanges of messages. Participants must keep in mind conversational context, and participants joining late can and often must acquire some of this context. In addition, some settings call for engaging in several conversations in parallel; in others, participants conduct temporarily limited sub-conversations to clarify a point. Existing actor code exhibits complex design patterns that get around the underlying limitations of the pure message-passing model.
**Approach**   These patterns suggest a number of elements involved in programming conversational computations. Translated into terms of language design, they call for two kinds of facilities: (1) one for sharing conversational context and (2) another one for organizing individual actors around on-going conversations and their contexts.
**Knowledge**   This paper presents Syndicate, a language designed to directly support the programming of conversing actors. Beyond message passing, it supplies (1) a dataspace, which allows actors to make public assertions, to withdraw them, and to query what other actors have asserted; and (2) the facet notation, which enables programmers to express individual actors as a reflection of the on-going conversations.
**Grounding**   A worked example introduces these concepts and illustrates conversational programming in Syndicate. A comparison with other research and industrial concurrent languages demonstrates the unique support Syndicate provides.
**Importance**   Syndicate advances concurrent actor programming with enhancements that address some observed limitations of the underlying model. While message-passing simplifies concurrent programming, it falls short in handling the complexities of actual computational conversations. By introducing a dataspace actor for sharing conversational context and the facet notation for organizing actors around ongoing conversations, Syndicate enables developers to naturally express and manage the nuanced interactions often required in concurrent systems. These innovations reduce the need for complex design patterns and provide unique support for building robust, context-aware concurrent applications.

# The Art, Science, and Engineering of Programming

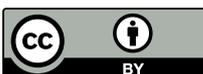

**Conversational Concurrency with Dataspaces and Facets**

## 1 Conversations and Concurrency

The actor model of computation presents concurrency as a conversation. At any point in time, each participating actor may send a message to another participating actor. In some cases the actor may wait for a response; in others, it continues to compute and send messages without waiting for a response. Organizing a concurrent computation as if it were a conversation is intuitive, and this approach avoids common pitfalls of other approaches to concurrency, e.g. overlapping access to shared memory.

A close look at real-world conversations suggests a picture that differs from the message-passing model. Let us peek at a table of dining, conversing attendees at the `<Programming>` conference. Experience suggests a number of observations:[1]

- The participants may engage in several conversations simultaneously. They may leave one conversation to join another, only to return to the first one later.
- Each such conversation is contextual. A newcomer should probably listen for a while to pick up (some of) the relevant context.
- As new topics are broached, a conversation may branch into several (sub-)conversations. Previous topics may be returned to—or not.
- The set of participants changes over time. Late arrivals join the table after the start of a conversation while others exit before its conclusion.
- Finally, these participants identify one another through a number of characteristics: their conference presentations, their association with particular research interests, etc. They may not even know each others' names. Indeed, a participant's identifying characteristic may vary from conversation to conversation.

At a minimum, this characterization of conversations suggests two insights. First, conversations are not free of shared state; participants keep conversational context in mind.[2] Second, participants juggle more than one conversation.

Syndicate, a marriage of the actor model with Linda's tuple spaces [16], supplies the tools for expressing both insights. Concretely, it includes two synergistic innovations. The first is the *dataspace model* of concurrency, a publish/subscribe variation of the actor model. The second is the language of *facets*, a notation for expressing the ever-changing behavior of an actor engaged in multiple conversations. These two innovations enable developers to program in the conversational style elaborated above, specifically to represent conversational context as shared program state and to set up actors to dynamically adapt their behaviors in the face of evolving conversations.

A *dataspace* is, roughly speaking, a special actor through which all communications are routed [14, 15, 33, 34]. It is a reification of the messaging substrate common to all actor systems. To share information, actors deposit facts in the dataspace. An actor wishing to receive certain messages or notifications about facts signals an interest to the dataspace. The dataspace sends updates to such actors as matching messages and facts come and go. When an actor terminates—gracefully or via an exception—the dataspace removes the actor's facts and interests.

---

[1] While Djikstra's dining philosophers illustrate how coordination avoids deadlock, the scenario of dining conference attendees paints a portrait of the actual kinds of conversations concurrent actors may have to conduct.

[2] Developers have noticed this point, and Tasharofi et al. [51] supply some empirical evidence.

The *facet* notation supports the automatic connection of an actor's control state with its published data; helps manage participation in multiple conversations; and implements automatic (de)multiplexing of incoming information. As the description of conversing researchers implies, all of these features are needed to formulate the code for an individual actor that participates in numerous, evolving conversations.

While past publications have presented the theoretical design of the dataspace actor model plus a theoretical argument for formulating actors with facets, this paper contributes a practical argument in support of programming dataspace actors with facets. Specifically, the goal is to demonstrate how easy it is to comprehend the workings of even complicated actors expressed with facets. The main body of the paper illustrates Syndicate with an example of multi-conversing actors (Sections 4 and 5). The example (Section 3) generalizes a well-known banking scenario to a simplistic stock market. The remainder of the paper compares Syndicate to similar, implemented models of computation, focusing on each model's support of a general conversational style of computing (Section 6). The remainder starts with an informal presentation of Syndicate's syntax and semantics (Section 2).

*Note.* Since Syndicate was first implemented in Racket, the paper comes with a rudimentary tour of this language in Appendix A; readers unfamiliar with Racket should read this section first.

## 2 Syndicate, the Hosted Language

Syndicate supplements Racket with two conceptual extensions. The first one is the dataspace-actor language (Section 2.1). The second is the *facet* notation. A facet is the fundamental behavioral unit of an individual actor (Sections 2.2 and 2.3). Together the two extensions enable the conversational style of actor programming described in the introduction (Section 2.4).—The section ends with a concise explanation of how this paper relates to previous publications about Syndicate's model.

*Note.* The examples color the forms of the two extensions red to distinguish them from base Racket forms, which are blue.

### 2.1 The Dataspace Model of Actors

A dataspace launches a group of actors and equips them with a means for information sharing. The pieces of `dataspace` expressions are function calls that spawn actors:

```
1  #lang syndicate
2
3  (define (main) (dataspace (atm-actor) (bank-client-actor)))
4  (define (atm-actor) (spawn atm-function))
5  (define (bank-client-actor) (spawn bank-client-function))
6      ___ code intentionally omitted ___
```

**Conversational Concurrency with Dataspaces and Facets**

The remainder of the module defines `atm-function` and `bank-client-function` via Racket's core language plus Syndicate-specific extensions.

To share a piece of information, an actor deposits it within the dataspace. If such a deposit satisfies the interests of some other actor, the dataspace notifies this actor about these pieces of information. Although the verbs "deposit" and "notifies" seem to imply imperative acts, a dataspace actually treats an actor as a (mathematical) function from an event notification to an action request. An event notification consists of positive and negative information concerning the publicly shared pieces of information. An action request calls on the dataspace to perform one (or more) of the following actions on the actor's behalf: (1) broadcast a message; (2) update its published information state; (3) spawn additional actors; and (4) clean up the published information in response to termination. Once the dataspace has executed these actions, it inspects the dataspace and decides which pieces of data should be routed to which actors.

To make this description concrete, an actor may publish the fact that some stock shares are available at the price of $100 by depositing an instance of the `price` struct

```
(price 100)
```

in the dataspace. When the price changes, the actor may withdraw `(price 100)` and introduce a new instance such as `(price 80)`. When the goods are no longer for sale, the actor retracts the instance without replacement.

Pieces of information in the dataspace are called *assertions* because they resemble facts in a Prolog database [19]. Assertions range over basic values—numbers, strings, etc.—as well as nested forms of immutable data structures, including lists and instances of user-defined structs.

An actor interested in public information creates an instance of the pre-defined `observe` Syndicate struct and deposits it. The `observe` constructor signals to the dataspace that the actor is interested in all assertions that match the *assertion pattern*. For example, a broker actor wishing to know that shares are offered at the price of $100 deposits

```
(observe (price 100))
```

A broker interested in any advertised price uses a wildcard pattern (⋆) instead of the concrete 100:

```
(observe (price ⋆))
```

When the dataspace notices this form of assertion, it notifies the actor each time the seller actor deposits *any* `price` assertion into the dataspace or withdraws it.

*Note.* An instance of the `observe` struct (in the dataspace) *is* an assertion. Hence, an actor providing a logging service may monitor activity with an assertion of the shape (`observe` (`observe` ___)).

Once deposited, an assertion is read-only and linked to the originating actor. Only an action by the owner may remove it, meaning either an explicit removal request or an implicit removal request because the actor terminated (one way or another).

The core of the dataspace is its routing logic. When an actor adds an `observe` to the dataspace, the routing logic notifies it

1. of any already-present assertions matching that interest; and
2. any time a matching assertion (dis)appears in the dataspace in the future.

Critically, the dataspace model interprets an assertion as the representation of some information in the conversation, meaning multiple identical assertions denote the same information as a single assertion of this kind. Hence, the routing mechanism presents an actor with a *set*-based view of the *bag* of assertions in the dataspace. Even if multiple actors make identical assertions, dataspaces hide this multiplicity. Conversely, when an actor is notified of the appearance of an assertion matching an interest, the notification denotes that the number of such assertions in the dataspace goes from zero to non-zero. Likewise, a disappearance notification means the number of assertions goes from non-zero to zero.

The implementation of dataspaces is re-programmed Racket as described in the preceding section. Its syntax seamlessly extends the core syntax. By contrast, the implementation of dataspaces in ECMAScript [24] is a library that requires programmers to write some boilerplate. Finally, an actor can, in principle, be expressed with any available linguistic construct from the base language, but supporting the conversational style spelled out in the introduction, demands additional linguistic support.

### 2.2 The Facets of Actor Behavior

From the perspective of a dataspace, an actor behaves as if it were a black box. It receives event notifications and responds with action requests as if it were a function. From the perspective of conversations taking place, this function composes atomic units of behavior; every piece of information in a notification adds or subtracts a unit of behavior. In the context of the Syndicate language, a unit of behavior is a *facet*. From the perspective of facets, an actor's behavior function corresponds to a dynamically evolving tree, reflecting the need to specify independent and nested behaviors.

The *facet language* supplies the syntax for expressing facets in a concise and comprehensible manner. In the Racket implementation of Syndicate, macros define new syntactic forms in the manner described in Section A.

Facets are created with the (`react`) form, which contains any number of *endpoints*:

- *assertion statements* represent requests to contribute publicly to the conversations;
- *event handlers* define reactions to changes in the conversation's public state; and
- *fields*, which hold the local and private state of an actor's take on the conversation.

Figure 1 displays the EBNF grammar for the facet notation. The red portions indicate names Syndicate introduces to Racket. The main entry point into this grammar is the actor production, which extends Racket's expressions.

**Spawning Actors**  The `spawn` form places a new actor into the dataspace. The body of the expression specifies the endpoints of its initial (root) facet.

**Starting Facets**  The `react` form creates a nested facet. When evaluated, it extends the actor's current behavior tree with a branch starting from the current node.

|  |  |  |  |
|---|---|---|---|
| actor expressions | | event specifications | |
| *actor* = (`spawn` *endpoint*[+]) | | = (`asserted` *pattern*) | |
| | (`react` *endpoint*[+]) | | | (`retracted` *pattern*) | |
| | (`stop` *name actor*[+]) | | | (`message` *pattern*) | |
| | (`send!` *actor*) | | | `start` | |
| | (`current-facet-id`) | | | `stop` | |
| endpoint specifications | | event patterns | |
| *endpoint* = *actor* | | *pattern* = (*constructor pattern*[*]) | |
| | (`field` {(*name actor*)}[+]) | | | *actor* | |
| | (`assert` *actor*) | | | `$`*var* | |
| | (`on` *event actor*[+]) | | | _ | |

■ **Figure 1** The Racket variant of the facet language

Listing 1 shows a concrete use of `react` (line 10) in the implementation of a simplistic bank account actor. The actor opens accounts at the request of clients and handles account balances, transactions, and so on, the details of which are elided. In response to a request for a new account, the bank actor creates a facet for this individual `client`. To recognize and manage such requests, the actor uses two kinds of endpoints: a reaction endpoint for the creation act and a field endpoint for managing the account balance.

■ **Listing 1** A simple bank actor

```
 1 (define (bank)
 2   (spawn
 3     (field (next-account-number 0))
 4     (define (get-next-account-number!)
 5       (define account-number (next-account-number))
 6       (next-account-number (add1 (next-account-number)))
 7       account-number)
 8     (on (asserted (create-account $client $initial-balance))
 9        (define account-number (get-next-account-number!))
10        (react
11          (field (current-balance initial-balance))
12          (assert (account-for client account-number))
13          (assert (balance account-number (current-balance)))
14          ⎯⎯ code intentionally omitted ⎯⎯))))
```

**Reactions** One kind of endpoint is the event handler. It combines two an event specification and a body of code to execute in response to a match of an incoming event with the specification. When the event handler becomes active, the actor constructs an `observe` assertion from the pattern and deposits it in the dataspace. Likewise the actor withdraws the interest when the event handler becomes inactive.

Listing 1 shows an event handler in the `bank` actor definition. Its event specification is `(asserted (create-account $client $initial-balance))`. It comes with a pattern that matches an event constructed with `create-account` and two values. The latter are bound to the two pattern variables, whose names start with `$` by convention. If an incoming event's data matches, the actor computes the next possible account number and starts a facet for this account using `react`.

Thus each distinct instance of a `create-account` assertion instantiates such a facet, branching from the root facet implicitly created by the `spawn` on line 2. As more clients open accounts, the width of the bank actor's facet tree grows.

**Fields: Connecting Public and Private State**   Facets typically maintain the key elements of their knowledge about a conversation in fields. Mechanically, a field manifests as a pair of read and write procedures. Calling the field with no arguments returns the current value (lines 5 and 13 of Listing 1) while a call with one argument sets the field to a new value (line 6, ditto). Pragmatically, these fields integrate pieces of information as they appear or disappear via notifications. In turn, a facet publishes parts of this knowledge in the dataspace. To support this continuous relationship between private and public state, the facet notation realizes a dataflow mechanism that automatically propagates data from fields to assertions. Concretely, when an assertion such as `balance` depends on the value of a field, here (`current-balance`), an update to the field causes the actor (1) to withdraw the related assertion, and (2) to deposit an assertion that reflects the updated field.

The scope of endpoints is their enclosing facet. Furthermore, facets install all endpoints dynamically. Hence pieces of behavior may be defined in separate functions or abstracted over into reusable parts. In the case of the `bank` actor, the root facet directly installs one `field` (line 3) and one reaction (line 8). The latter installs `current-balance`, which leads to a distinct cell per instantiation of the enclosing facet (line 10), and contributes two assertions to the dataspace, one of which is integrated with `current-balance`.

**Stopping Facets**   The `stop` form initiates the shutdown of the specified facet and all of its children in the behavior tree. Designating the facet to terminate requires invoking the `current-facet-id` procedure, which returns a value that uniquely identifies the enclosing facet. As the shutdown process completes, the actor also de-allocates the facet's fields and issues a withdrawal request for all assertions associated with this subtree. The second part of `stop` represents its *continuation* behaviors. They execute after the shutdown process completes, in place of the just-terminated facet, grafting replacement facets onto the root of the sub-tree.

**Synthetic Events**   The facet language comes with endpoints for reacting to the launch or the tear-down of a facet, respectively: `start` and `stop`. While a `start`-event handler helps initialize a facet and acquire resources, a `stop`-event handler performs clean-up actions and releases resources. The `stop` handlers run only during orderly shutdown; a facet termination due to an exception does *not* trigger a `stop` event. Since

**Conversational Concurrency with Dataspaces and Facets**

the very nature of an uncaught exception implies that any and all state invariants may be violated, triggering a `stop` event would likely yield undesired behaviors.[3]

**Messages** In addition to assertions, dataspaces also support message broadcasting. The `send!` form broadcasts a message through the dataspace. Actors that have deposited a message interest—(`message` *p*)—receive messages that match p.

## 2.3 Abstractions over Actor Behavior

Facets make up the fundamental building blocks of actors that conduct several conversations concurrently. Their syntax is simple and easy to comprehend. However, programmers using facets to implement assertion-based protocols quickly notice repeating higher-level patterns deserving direct syntactic representation.

```
1 (during pattern
2    during-body
3    ...)
```
(*where pattern′ instantiates every pattern variable in pattern with its concrete value.*)

$\stackrel{\text{def}}{\equiv}$

```
1 (on (asserted pattern)
2    (react
3       (define during-body
4          (current-facet-id))
5       (on (retracted pattern′)
6          (stop during-body))
7       during-body ...))
```

**Figure 2** The `during` form

Two important derived forms show up in several actors in the following sections: `during` and `state-machine`. The `during` form implements the common temporal pattern of creating a facet in response to the appearance of an assertion and tearing it down again in response to its disappearance. Figure 2 displays the implementation of `during` as a simple notational abbreviation over the base notation.

Likewise, the `state-machine` form sets up a group of related behaviors, of which only one is active at a time. On some occasions, the transition among these behaviors is just a linear chain; on others, the transitions make up a cyclic graph. Appendix C provides the definition of the `state-machine` form.

The `during` and `state-machine` forms are not a part of Syndicate's core. Instead, they are user-level abstractions that distill common programming patterns using the language-extension mechanism inherited from Racket. As programmers gather more experience with these linguistic forms, they may find improvements or other syntactic patterns that deserve the same status as `during` and `state-machine`. Critically, functional and notational abstractions over the facet notation seamlessly compose (in the Racket-based Syndicate version) with other abstractions and primitive forms.

---

[3] This explains the language-level support for exceptions in Syndicate.—Determining the exception-safety [1] of specific code within an actor typically requires reasoning about subtle dependencies between the state invariant and cleanup procedures. Due to the difficulty of correctly understanding every such case, we recommend the *fail fast* philosophy of Erlang/OTP [4] to Syndicate programmers. This guideline says "when in doubt, crash," allowing other actors in the system to detect the fault and take appropriate action.

## 2.4 Conversational Style

Finally, let us connect the linguistic constructs to the proposed conversational style:

- Engaging in a new conversation is as simple as starting a facet. Likewise, stopping a facet disengages from a conversation.
- The assertions in the dataspace represent conversational context. The `during` form connects this context to the actor's behavior on a temporal basis.
- Facet trees mirror the branching and nested nature of conversations.
- When a new actor joins a conversation, its event handlers seamlessly match both the existing context and ongoing communications. Likewise, when it terminates, the dataspace removes its assertions to maintain an accurate picture.
- Data-driven routing in the dataspace enables domain-specific notions of identity, which may vary from conversation to conversation.

## 2.5 Relationship to Prior Work

The aim of this paper is to argue the *pragmatics* of Syndicate, i.e. to demonstrate the usage of the various linguistic mechanisms and forms in context. By contrast, prior publications on Syndicate [13, 14, 15, 32, 33] primarily focused on the semantics of the dataspace model, its type system, and the verification of properties.

Garnock-Jones et al. [33] introduced dataspaces as a means of coordinating actors via the routing of assertions for inter-actor communication. Their presentation treated each actor in the system as a function (programmed in core Racket). Experience with dataspace programming revealed the need for additional linguistic support in the form of facets [14, 32]. Caldwell et al. [13, 14, 15] explored type system support for dataspace and facet programming, with an eye toward lightweight verification.

## 2.6 Implementations

Syndicate has been implemented and integrated with several programming languages besides Racket. A Rust implementation enjoys roughly the same status as the Racket implementation, due to the language's Racket-like (and Racket-inspired) macro system. Concretely, Syndicate code in Rust looks and feels like extended Rust programs, similar to the Syndicate-Racket programs in this section. The macro systems of both languages also cooperate with IDEs to some extent, meaning Syndicate programs come with some IDE support. By contrast, the implementations of Syndicate in Python, Smalltalk, and Java exist only in the form of libraries, meaning they lack syntactic support and thus IDE support. None of the implementations, besides the Racket one, realize the type system of Caldwell et al. [14, 15], but see Appendix E for the JavaScript/TypeScript implementation, including some sample code. Distributed components written in the various implementations may interoperate using a common Syndicate network protocol.

Conversational Concurrency with Dataspaces and Facets

## 3 The Scenario

Programming in Syndicate substantially differs from programming in conventional actor languages, a difference that is best explained with an illustrative, extensible example. The chosen example is a play on the banker-client tradition of this area. Concretely, the goal is to simulate the concurrency aspect of buying stock from a seller through a broker. The scenario involves some $n$ buyers, a single seller, a broker, and a banker. Buyers come and go, and they purchase shares of a single stock via the broker in an unpredictable order. Each stock order specifies the number of shares to acquire and the maximum price per share that the buyer is willing to pay. At this point, the broker takes over to fulfill these purchasing requests, subject to these conditions:

**timing** The broker, seller, and bank work only during trading periods.

**funding** Each buyer has a bank account.
The broker proceeds with an order only if it can withdraw a sufficient amount from the buyer's account to complete the purchase at the order-specified price.

**pricing** The transaction may complete only if shares are available at a price compatible with the buyer's request.

**canceling** The buyer may cancel an order until the broker has settled the purchase.

Implementing this setup requires $n+4$ actors communicating via one dataspace:

$n$ **buyers** making, monitoring, and optionally canceling orders at will;

**the broker** working to complete orders during the trading period;

**the bank** maintaining each buyer's account; and

**the seller** setting the price of shares and selling them.

**the clock** announcing the opening and closing of trading periods.

Fulfilling orders requires several kinds of interactions among these actors:

**ordering** The buyer initiates an order with the broker by specifying the desired number of shares and maximum price per share. After returning all unused funds to the buyer's bank account, the broker responds with a confirmation that the order is
- fulfilled successfully;
- rejected due to insufficient funds in the buyer's bank account;
- canceled by the buyer; or
- unable to complete at the specified price.

**banking** Each buyer has an account with the bank. The broker may withdraw funds from, or deposit funds into, a buyer's account.

**pricing** The seller advertises the current share price.

**purchasing** Only the broker can purchase shares from the seller.

**Conversational Properties** With respect to the conversational style described in the introduction, the scenario exhibits the following characteristics:

⇒, ⋖: For each order, the broker conducts several conversations in order: with the buyer, the bank, and the seller, before returning to the conversation with the buyer.

- 🗄: Conversations with the bank, the broker, and the seller are contextual; they depend on the cycle of trading periods.
- 👥: For simplicity, the example uses a single broker and seller, and only buyers enter or leave at any time.
- ∞: The scenario uses several identifying characteristics: the buyer comes with a bank account number as well as with order numbers for the broker. Indeed, all transactions use IDs to distinguish them from each other.

## 4 Syndicate, by Example

The combination of dataspaces with facet-oriented actors can directly express the conversational style described in the introduction in a comprehensible manner.

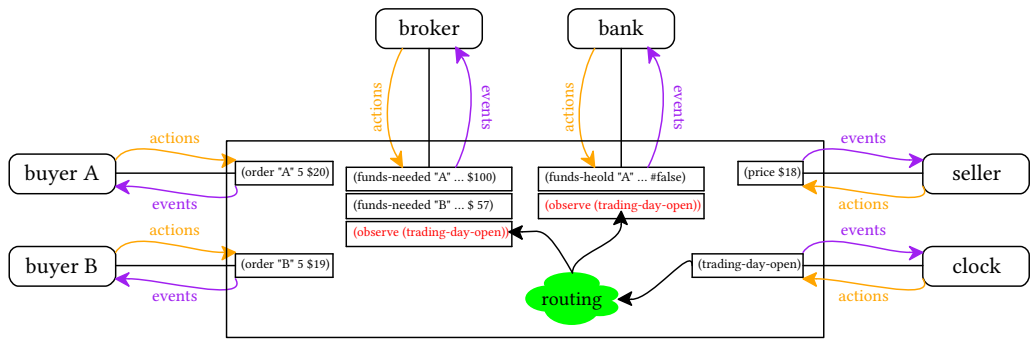

**Figure 3** Visualizing select elements of the running market program

### 4.1 The Architecture

Figure 3 depicts some elements from a potential run of the program. The large rectangle in the center represents the dataspace, a container of its connected actors' *assertions* (black text) and *interests* (red text)—depicted as small rectangles inside the dataspace. The rounded rectangles surrounding the dataspace represent actors.

An actor is related to its assertions and interests in three different ways. First, it places assertions and interests into the dataspace (orange arrow and label). Second, it owns its assertions and interest (black line). Third, it receives events (purple arrow) when the dataspace notices a change to assertions relative to an actor's interests.

The green cloud inside the dataspace represents the routing mechanism. When an assertion (dis)appears, the dataspace informs all actors with matching interests of this event. Since assertions may (dis)appear simultaneously, an actor may receive multiple pieces of event information at once.

Each assertion, message, and interest belongs to the communication protocol.

**Conversational Concurrency with Dataspaces and Facets**

```
A  protocol assertion is one of:                          An  id is unique value.
  (trading-day-open)
  (order id account-number no-of-shares price/share)      An  answer is one of:
  (order-result order answer)                              - 'canceled
  (funds-needed id account-number amount)                  - 'insufficient-funds
  (funds-held id account-number amount Boolean)            - 'no-price-match
  (price price/share)                                      - 'fulfilled
  (purchase-request id no-of-shares price/share)
  (purchase-result id Boolean)
```

**Figure 4** The protocol for simulating a simple market

### 4.2 The Protocol

Figure 4 specifies the protocol in terms of the shape of the Racket structs that show up in the dataspace. The various structs pertain to the following elements of the protocol:

**timing** The clock publishes (`trading-day-open`) to signal the beginning of a trading day; it withdraws this assertion at the end of the day.

**ordering** To request a purchase of $N$ shares for a maximum per share price of $P$, buyer $i$ publishes (`order id A N P`), where $A$ is the buyer's bank account number and *id* is any value that uniquely identifies the `order`. To request cancellation of the order, the buyer withdraws its `order`.—The broker responds to the request by publishing its result, (`order-result O answer`), where $O$ is the `order` itself.

**banking** The broker starts working on a buyer's order with a request to the bank by publishing (`funds-needed id A` $(N \times P)$). Here *id* is any value that uniquely identifies the request. The bank responds with an assertion indicating the success or failure of acquisition: (`funds-held id A` $(N \times P)$ `Boolean`). (This description elides one intermediate step; Section 4.5 provides the full details.) If the order is canceled or fails to attract the seller's interest, the broker can deposit unused funds back via a similar exchange.

**pricing** The seller publishes the current price of a share $S$ as (`price S`).

**purchasing** The broker asserts (`purchase-request id N P`) to purchase shares from the seller. The seller responds with (`purchase-result id Boolean`).

### 4.3 Actor Internals

Facet-based actors mirror dataspace protocols with their internal structure. The behavior of each actor is a tree of facets that grows and shrinks in response to conversation requests in the datasapce.

The diagram in Figure 5 depicts the control state of the broker at a particular moment (blue box). The root of the tree is a node that activates or deactivates the entire tree of behaviors depending on whether trading is open. When trading is open, every appearance of an `order` assertion causes the creation of a branch in the tree. Branches evolve independently but sequentially. Here the first step of working on an order calls for acquiring the necessary funds, meaning the branch grows by a facet.

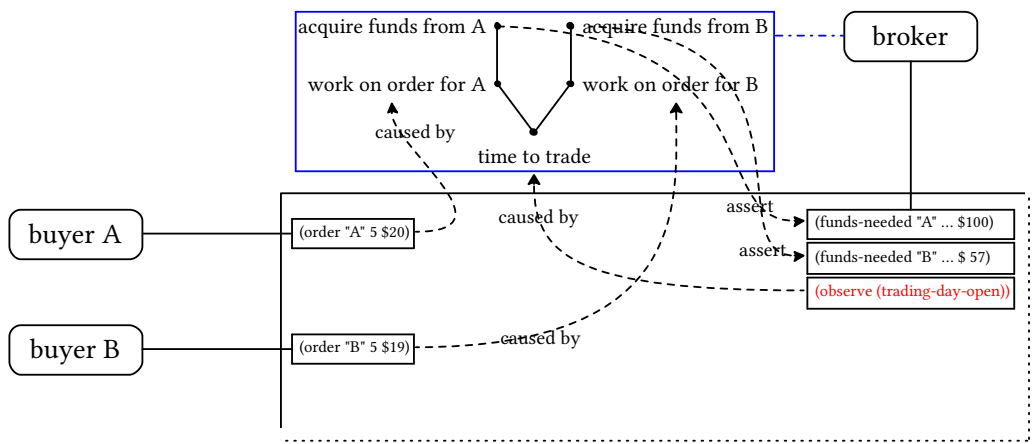

**Figure 5** Visualizing the internal state of the broker actor

In the diagram, the dashed lines indicate the connections between each facet of the actor's behavior tree on one hand and assertions or interests (in the dataspace) on the other. The connection between the broker's interest in the clock's assertions is depicted accurately; the one between orders and their internal nodes is just a sketch. In reverse, new nodes in the tree place their own assertions into the dataspace, such as `funds-needed` when working on an `order`.

### 4.4 The Implementation

The `main` function initializes the dataspace program for the market simulation:

```
1 (define (main)
2  (dataspace
3   (clock ──) (bank ──) (seller ──) (broker ──) (buyer-factory ──)))
```

It creates the four core actors by calling appropriate functions plus an actor that spawns as many buyer actors as desired.

**The Seller** Figure 6 defines the function that creates the seller, the simplest of all actors. Its behavior is contextual: it operates when trading is open but is idle otherwise. The `during` form enables expressing such contextual behavior as directly as possible. It activates some behavior when the dataspace contains the designated assertion and deactivates it when this assertion disappears. In this case, the rest of the seller's behavior is contingent upon the presence of the `trading-day-open` assertion.

When the trading day is open, the seller performs two tasks. First, it publishes the asking price of shares as an assertion. Second, it responds to requests to purchase shares. Perhaps surprisingly, it also controls this second behavior with a `during` form. In this case, the `during` matches on an assertion pattern. Specifically, it matches instances of the `purchase-request` struct and binds its pieces to the `id`, `count`, and `offered` variables, respectively. With such a pattern, the inner behavior activates once

```
1 (define (seller desired)
2   (spawn
3     (during (trading-day-open)
4       (assert (price desired))
5       (during (purchase-request $id $count $offered)
6         (assert (purchase-result id (>= offered desired)))))))
```

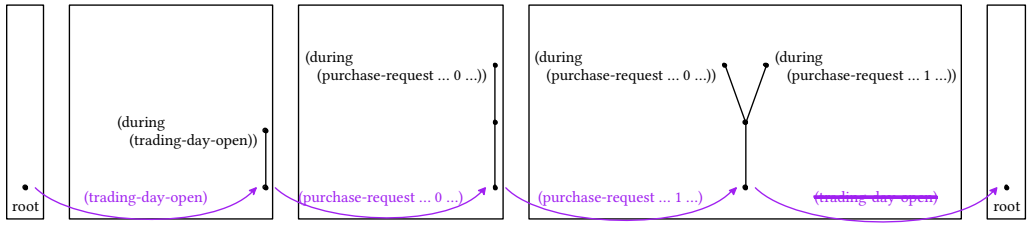

**Figure 6** Top: definition of the seller. Bottom: evolution of the actor's behavior tree

per *distinct assertion* matching the pattern. Thus, each `purchase-request` receives a response based on the offered and desired prices.

An implementation based on `during` provides garbage-collection-like clean-up functionality. In particular, the `during` form monitors the *relevancy* of its enclosed behavior to the current conversation. When the broker withdraws its `purchase-request` assertion upon receiving a response, the `during` in `seller` deactivates.

The bottom of Figure 6 illustrates the evolution of the seller's behavior tree as other program components publish assertions. Each box represents a point in time; the purple arrows between them indicate new or removed assertions triggering a reaction from the actor. Every tree has a single root that monitors the trading day. When the trading day opens, the behavior of the corresponding `during` extends this root and begins monitoring purchase requests. Each purchase request gives rise to a new branch in the tree. When the trading day ends, the behavior tree is pruned back to its root.

**The Broker** Figure 7 defines the function for launching the broker. Similar to the seller actor, its behavior is contingent upon the presence of a (`trading-day-open`) assertion. During such periods, the broker works on fulfilling buyers' `orders`. The (`on (asserted __) __`) reaction endpoint extend the actor's behavior tree for each `order` assertion. The `work-on-one-order` function implements the first step of this process. The actor provides the helper function with the order and a representation of the particular control context in the actor that corresponds to this particular order. The curried `stop-with!` procedure uses this context to create a function, which, when applied to an answer, prunes the sub-tree of behaviors above the given context. Section 4.5 presents the implementation of `stop-with!`.

The bottom of the figure shows the evolution of the actor's behavior tree. The tree grows and shrinks like the one of the seller due to their analogous uses of `during`. For each order, the `work-on-one-order` function further extends the tree. When it completes an order, it issues a `order-result` assertion (orange arrow) and shuts down its branch of the behavior tree.

```
1  (define (broker)
2    (spawn
3      (during (trading-day-open)
4        (on (asserted (order $id $account $amt $desired-price))
5          (react
6            (define the-order (order id account amt desired-price))
7            (define cfid (current-facet-id))
8            (work-on-one-order the-order (stop-with! the-order cfid))))))))
9
10 (define ((stop-with! the-order order-context) answer) ___)
```

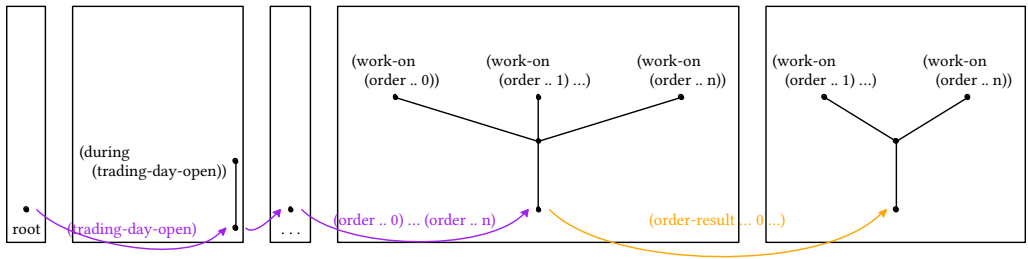

**Figure 7** Top: definition of the broker. Bottom: evolution of the actor's behavior tree

Processing an order via `work-on-one-order` proceeds in two stages: (1) acquiring the funds necessary to cover the transaction from the buyer's bank account; and (2) complete a fully-funded order. The interaction thus takes the form of a `state-machine` stepping through two states:

```
1  (define (work-on-one-order the-order stop-with!)
2    (define potential-cost (max-order-cost the-order))
3    (state-machine acquire-funds
4     (request (define (on-funded) (goto funded))
5              (request-funds the-order potential-cost on-funded stop-with!))
6     (funded (finish-funded the-order potential-cost stop-with!))))
```

The two states realize exactly the two stages of the order-processing conversation:

**request** requesting the funds from the bank and interpreting the response; and

**funded** finish processing the order, when it is funded.

The reader must keep in mind that, at any point during these conversations with the bank and seller, respectively, the buyer may cancel its order.

Figure 8 presents `request-funds`, a function that requests the necessary funds. It first publishes a `funds-needed` assertion and registers three different handlers to react to three kinds of events. These event handlers augment the behavior of their containing `during` facet. Hence, any activation of `stop-with!` in these handlers tears down the entire `during` behavior. The first two event handlers react to the two possible responses to the just-published assertion:

**#true** If the bank can transfer a sufficient amount, `request-funds` initiates a transition of the `acquire-funds` state-machine to its second state.

**#false** Otherwise, the broker stops the processing of this order.

```
1 (define (request-funds the-order potential-cost on-funded stop-with!)
2   (define account (order-account the-order))
3   (assert (funds-needed the-order account potential-cost))
4   (on (asserted (funds-held the-order account potential-cost #true))
5       (on-funded))
6   (on (asserted (funds-held the-order account potential-cost #false))
7       (stop-with! 'insufficient-funds))
8   (on (retracted the-order)
9       (stop-with! 'canceled)))
```

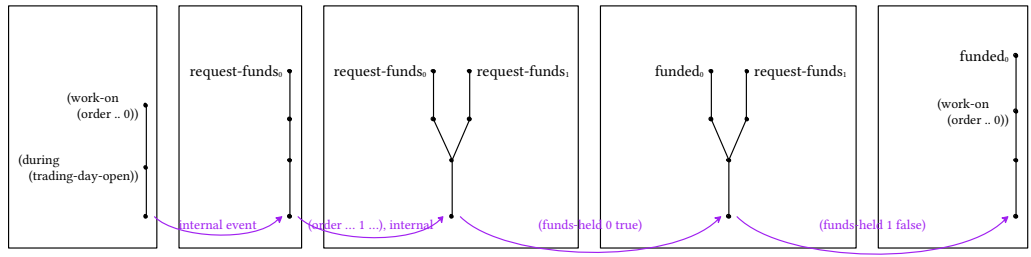

**Figure 8** The broker acquires the necessary funds

In case the actor receives either one of these two events, an activation of the third handler can no longer happen. Otherwise, if the buyer cancels the order with a retraction of its corresponding assertion, the third event handler invokes the `stop-with!` continuation with a confirmation of the cancellation and, because no funds have been withheld yet, does not need to return any.

The bottom of Figure 8 depicts the manner in which the `acquire-funds` state machine augments the actor's behavior tree. As mentioned above, the tree adds a branch for each active order. The state machine created by the `work-on-one-order` procedure extends each of these branches independently. As information for each order arrives at the broker, such as a `funds-held` assertion, only the state machine along that particular branch of the tree evolves.

The finalization of a buyer's order also calls for two stages: (1) requesting the current price from the seller; and (2) completing the purchase with the seller if the price fits the buyer's expectations. To simplify the presentation, assume that such purchase requests always succeed. At both stages, the behavior also terminates the outer conversation with the buyer, as needed.

Figure 9 presents the function that implements this behavior. The function also uses a state machine to implement the staged behavior:

**request-price** still comes with a handler for order cancellation. It differs from the one in `request-funds`, because the acquired funds must be returned.

**complete-purchase** does not include a handler. Once the seller meets the desired price of the buyer, the broker commits to purchasing the desired number of shares. Consequently, any request from the buyer to cancel the order will be ignored.

Notice how the second state takes a parameter for the `actual` price of shares, as determined in the first state. The `goto` passes the value provided on line 4 to the code on line 7 in the expected manner.

```
1  (define (finish-funded the-order held-funds stop-with!)
2    (state-machine purchase-progress
3      (request
4        (define (on-match actual) (goto purchase actual))
5        (request-price the-order held-funds on-match stop-with!))
6      ((purchase actual)
7        (complete-purchase the-order held-funds actual stop-with!))))
8
9  (define (request-price the-order held-funds on-match stop-with!)
10   (match-define (order _ account _ desired-price) the-order)
11   (on (asserted (price $actual))
12       (if (<= actual desired)
13           (on-match actual-price)
14           (stop-with! 'no-price-match)))
15   (on (retracted the-order)
16       (return->stop account held-funds 'canceled stop-with!)))
17
18 (define (complete-purchase the-order held-funds actual stop-with!)
19   (match-define (order _ account desired-no-shares _) the-order)
20   (assert (purchase-request the-order desired-no-shares actual))
21   (on (asserted (purchase-response the-order #true))
22       (define left-over (- held-funds (* desired-no-shares actual)))
23       (return->stop account left-over 'fulfilled stop-with!)))
24
25 #; {AccountId PositiveReal Symbol (-> Symbol Empty) -> Empty}
26 (define (return->stop account held-funds reason stop-with!)
27   (deposit! account held-funds)
28   (stop-with! reason))
```

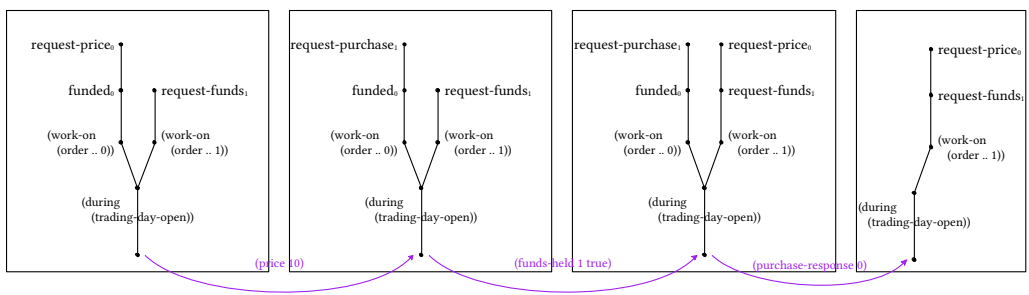

**Figure 9** The broker completes the order

The bottom of Figure 9 illustrates the effect of `finish-funded` on the actor's behavior tree. Much like the `state-machine` from `work-on-one-order`, `purchase-progress` evolves independently along each branch of the tree. When a purchase completes, the entire sub-tree for this order is removed.

### 4.5  Pausing and Resuming Work on Orders, Terminating Conversations

According to the specification, the broker must stop work when the trading day closes and pick up where it left off when trading reopens. The broker realizes the first part of this specification by shutting down the `during` facet. Since the termination of a facet would remove all associated conversational context, the broker spawns an auxiliary actor, `wallet`, which acts as an intermediary between the broker and bank actors.

Concretely, the broker must remember its private (control) state concerning the funding status of every open order. To this end, it communicates with the wallet actor concerning the `funds-needed` and `funds-held` assertions in `request-funds`. The wallet relays the initial request and response to the bank. Unlike the broker, the wallet maintains its public state across trading days. It can thus answer a request to fund an already-funded order without forwarding it to the bank. The actor withdraws funding assertions when it notices an `order-result` assertion for the corresponding order.

■ **Listing 2**   Shutting down an order behavior
```
1  (define (stop-with! the-order order-context)
2    (field (termination-reason #false))
3    (lambda (answer)
4      (unless (termination-reason)
5        (termination-reason answer)
6        (stop order-context
7              (spawn (order-result-caching the-order answer))))))
```

The broker employs the same technique in order to preserve the `order-result` assertions when it stops the facet for a single order. The auxiliary actor it spawns places an assertion with the answer for one order in the dataspace; see Listing 2. The elided `order-result-caching` function waits for the buyer to confirm receipt and then shuts down the actor, because the result assertion is no longer needed.

Dataspace communication concerns changes in assertions and a single action by an actor may induce numerous changes, introducing some new ones while removing others. As such, an actor may process a notification that activates a constellation of different event handlers across its facets. For example, the broker may simultaneously receive the confirmation for a stock purchase and a cancellation request from the buyer. Given the contradictory nature of those two events, some consideration must be given to how the actor should handle them to produce the correct response.

The solution employed by the broker actor is a combination of explicit state and a default order for event processing. The implementation uses a mutable `field` to ensure each order receives only one answer. Line 4 reads the `termination-reason` field while line 5 sets it. Thus, the first call to `stop-with!` determines the result of the order. An event may trigger multiple handlers that call `stop-with!`. To make the outcome predictable to the programmer, Syndicate runs the handlers in the same order in which they were registered.

## 5 Extending the Example

The communication protocol in the scenario of the preceding section could mostly have used point-to-point messaging instead of an assertion-based mechanism. Consider, though, how much such a direct-messaging protocol would have to change if the scenario involved numerous brokers, each competing for business from buyers via fee advertisements, and numerous sellers, competing for business from brokers via price advertisements. Generalizing from the simple scenario to this complex one is *straightforward* with a proper conversational communication mechanism, such as dataspaces and facets.

This section explains how to adapt the code from the preceding section with minimal adjustments. Critically, in a facet-based program, programmers may modify each basic (code for a) node in the behavior tree without (necessarily) affecting others.

Revising the implementation for the general case requires four steps:

1. introducing identifying values for brokers and sellers;
2. adding such identifiers to the various structs of the protocol;
3. updating the behavior of the broker to select among sellers; and
4. applying a similar change to the buyer for selecting brokers.

**Names** Let's consider one illustrative example of updating the protocol. Rather than asserting just a `price`, a seller actor shares (`price` *name price/share*). Then, when making a `purchase-request`, a broker actor includes the seller's `name` it wishes to purchase from. The other changes to the protocol follow the same pattern.

**Listing 3** Revising the interaction between broker and seller(s)

```
1 (define (finish-funded the-order held-funds stop-with!)
2   (state-machine purchase-progress
3    (select
4     (define (go-finish seller actual) (goto purchase seller actual))
5     (select-seller the-order go-finish stop-with!))
6    ((purchase seller actual)
7     (define (go-select) (goto select))
8     (complete-purchase the-order seller actual go-select stop-with!)))))
```

**The Broker** Listing 3 displays how the broker's `finish-funded` function must change. Instead of just requesting a price, the first state of its state-machine must `select` a seller before moving to the `purchase` state.

For the selection of the seller, the revised broker queries the dataspace as if it were a database. Using `define/query-hash`, it collects all available `price` ads for a period of time before choosing the best offer. This derived syntax collects entries in a hash table mapping sellers to their offered prices as their assertions show up in the dataspace. If a seller removes an assertion while this form is active, the corresponding entry is removed from the hash table. See Listing 4 for the definition of `select-seller`.

**Listing 4** Selecting among available sellers

```
1  (define (select-seller the-order held-funds go-finish stop-with!)
2    (match-define (order _ _ account _ desired) the-order)
3    (define/query-hash sellers (price $seller $actual) seller actual)
4    (on (retracted the-order)
5        (return->stop account held-funds 'canceled stop-with!))
6    (on (timeout WAIT-PERIOD)
7        (decide sellers held-funds desired go-finish)))
8
9  (define (decide sellers held-funds desired go-finish)
10   (cond
11     ((none? sellers)
12      (return->stop account held-funds 'no-price-match stop-with!))
13     (else
14      (match-define (best-seller actual)
15        (pick-best-seller-and-price sellers))
16      (if (<= actual desired)
17          (go-finish best-seller actual)
18          (return->stop account held-funds 'no-price-match stop-with!)))))
```

After the specified time period is over, the `timeout` handler runs the `decide` function. If no sellers are available, the broker informs the buyer with `'no-price-match` and returns the funds held. Otherwise, the broker actor picks a `seller` offering the lowest `actual` price. Assuming the `actual` price is lower than the buyer's `desired` one, the actor completes the purchase.

The `(on (timeout ___))` form is an example of an interaction with a dataspace *driver* actor. A driver actor provides an interface in terms of dataspace assertions and messages for an external resource or device. For example, Syndicate comes with a timer driver which introduces the notions of absolute and relative time to dataspace communications. The Racket implementation of Syndicate includes driver actors for network communication, file system access, and so on, while the JavaScript implementation includes a driver for accessing and updating the DOM. Moreover, Syndicate features an API allowing any programmer to create new drivers; the details of the API are beyond the scope of this presentation. The `timeout` form is a notational shorthand provided by Syndicate to communicate with the timer driver actor. The form creates an event handler that uses information from the driver to activate after a given number of milliseconds. The Syndicate implementation come with an array of driver actors for common needs, including file and network I/O, GUI interactions, etc. Programmers may of course implement their own driver actors as needed.

Listing 5 shows the last code adaptation. Concretely, this revision of the `complete-purchase` function changes the original from Figure 9 in mostly one piece, namely, when the seller withdraws its offer before the purchase completes. In that event, the broker goes back to the `select` state.

**Buyer Behavior** The other significant change to the program updates the behavior of a buyer to choose among available brokers. The modifications correspond to the revised

**Listing 5** Completing the purchase

```
1 (define (complete-purchase the-order seller actual go-back stop-with!)
2   (match-define (order _ _ account desired-no-shares desired-amount)
3     the-order)
4   (assert (purchase-request the-order seller desired-no-shares actual))
5   (on (asserted (purchase-response the-order seller #true))
6       (define left-over (- held-funds (* desired-no-shares actual) FEE))
7       (return->stop account left-over 'canceled stop-with!))
8   (on (retracted (price seller _))
9       (go-back)))
```

definition of the `finish-funded` procedure. The buyer first collects information about available brokers before selecting a preferred one for its order.

**Exercise** The reader may wish to contemplate the changes needed to adapt a traditional actor implementation of the market simulation to this generalized scenario.

## 6 Comparisons to Related Work

The three preceding sections make the case that the combination of the dataspace model of actors and facet notation provides an expressive substrate for tackling the conversational form of concurrent programming. This section assesses several alternative concurrent models as implemented in programming languages with respect to conversational programming. The assessment helps illustrate the differences among these approaches and offers a potential guide to future research.

### 6.1 Conversational Style

Before diving into a comparison, let us revisit the elements of conversational style and the manner in which Syndicate achieves it within the context of the example:

⇉: The broker engages in many conversations simultaneously. For each additional order, its behavior tree grows in width.

🗄: The broker engages in conversations for the trading day. The `during` form makes this direct connection to relevant conversational context obvious.

⤳: In the course of working on each order, the broker conducts conversations with the buyer, bank, and seller before returning to the buyer. For each such nested conversation, its behavior tree grows in depth.

🥸: The program uses two notions of identity: transaction IDs and bank account numbers. The actors include identity information only when needed. Thus, the wallet may act as a transparent intermediary between the broker and bank.

👥: The implementation of the broker actor allows buyers to come and go. When the scenario features numerous brokers and sellers, it suffices to simply enrich the data definition of identification information to designate a specific broker or seller.

**Conversational Concurrency with Dataspaces and Facets**

## 6.2 Concurrency Models and Their Implementations

Concurrency models come in a wide variety of flavors. Implementations of each differ almost as much. This section mostly uses five of the former to group descriptions of the latter; it also covers three hybrid models with one implementation each because they come quite close to supporting conversational concurrency properly.

**Actors** Akka [3] and Erlang/OTP [4] implement the actor model [2, 22]. The primary basis of communication and coordination between concurrent components is point-to-point, asynchronous message passing. To send a message, an actor must have a reference uniquely identifying the intended recipient. All messages received by an individual actor go into a single queue ("mailbox") that the actor processes one-by-one.

**Active Objects** Active Objects [8, 12, 47, 48] extend actor-style point-to-point messaging with concepts of object-oriented programming. An active object is an actor implemented as an object. Rather than sending and receiving messages, active objects tend to communicate in terms of asynchronous method invocations and their results. Researchers have used the method/message abstraction boundary as the basis for distributing control and state within an individual active object.

**Channels** Channels provide a concrete representation of the bidirectional flow of messages between concurrent parties. Available in both synchronous and asynchronous variety, message-passing over channels has gained significant industrial adoption (Go [36]) and seen significant academic interest in the $\pi$-calculus [40] and its derivatives. CML [46] demonstrates how channel-based programs can weave together a variety of different event sources to synthesize behavior.

**Multithreading with Shared Memory** Historically, many concurrent programs have relied on memory shared among multiple threads. These systems come with a variety of synchronization mechanisms to coordinate computations. The Java [37] language typifies this style. There have also been efforts to provide structure to the concurrent behavior of threads, such as the fork/join style provided by Cilk [31].

**Tuplespaces** Tuplespaces resemble a blackboard system, where one process deposits messages in a shared space that are read and removed by other processes. The Linda language [16, 35] makes tuplespaces available for the coordination of concurrent threads. The LIME language [44] extends Linda with *reactions*, handler functions that run when matching tuples appear but, notably, not as they disappear.

**Hybrids** *Fact Spaces and CRIME*. The *fact spaces model* [43] and its implementation, dubbed CRIME [42], shares many traits with the dataspace model. System components register facts in a Prolog-style database. The developer then writes a logic program that analyzes the facts with the capability of registering new facts and activating application callbacks. By recording the connections between facts and callbacks, CRIME applications may efficiently react to new and disappearing facts.

*Multi-agent Systems and Jason.* Jason [11] and its extensions [9, 10] to the AgentSpeak(L) language [45] is a framework for specifying multi-agent simulation and planning systems [50]. It resembles Crime in that updates to a Prolog-style database of facts, dubbed beliefs, trigger events for agents in the system. Each agent performs actions in pursuit of some goal state(s), which evolve based on current beliefs and executed actions. Each agent may concurrently pursue several different goal states.

*E, AmbientTalk.* The E language [39] extends the concept of an active object to a *vat* of colocated objects. Each vat runs its own event loop, dispatching incoming messages (method-invocations) to the contained objects one at a time. The AmbientTalk language [53, 54] shares the core design of E and introduces concepts for mobile computing and its challenges. Chief among these mechanisms are publish/subscribe-based service discovery and tuplespace-based state repositories [49].

*Sparrow.* The Sparrow DSL [5] takes inspiration from join patterns [7, 30] and extends Erlang-style Actors with a language of patterns and reactions. A pattern may specify a combination of messages arriving in the actor's mailbox, together or over time, and may be subject to a number of constraints or side conditions. Reactions are arbitrary application callbacks that are dynamically attached and removed from patterns. Each time a pattern matches, each of its attached reactions execute.

### 6.3 Results

A comparison along the dimensions of conversational concurrency must focus on directly expressed or similar capabilities, while it must ignore that each language may encode an implementation of another. It must allow the use of localized patterns or library abstractions. The key is that the use of these must not force the creator of one module to force the developers of others to change how they code [25].

| Paradigm (Implementations) | = | 🗄 | < | 👥 | ∞ |
|---|---|---|---|---|---|
| Actors (Erlang/OTP [4], Akka [3]) | | | ☆ | ☆ | ☆ |
| Active Objects (Encore [12], JCoBox [48]) | ★ | | ☆ | ☆ | ☆ |
| Channel-based (Go [36], CML [46]) | ☆ | ☆ | ☆ | ☆ | ☆ |
| Shared Memory (Java [37], Cilk [31]) | ☆ | ☆ | ☆ | ★ | |
| Tuplespaces (Linda [16], Lime [44]) | | ☆ | ☆ | ★ | ★ |
| Hybrid 1 (Fact Spaces [43]) | | ★ | | ★ | ★ |
| Hybrid 2 (Jason [11]) | ★ | ★ | ☆ | ★ | ★ |
| Hybrid 3 (E [39], AmbientTalk [54]) | ★ | ★ | ☆ | ★ | ★ |
| Hybrid 4 (Sparrow [5]) | ☆ | ☆ | ☆ | ☆ | ☆ |

▪ **Figure 10** Analysis of concurrent languages

Figure 10 shows the authors' assessment of each model's capabilities in terms of implementation. The marks indicate support of an aspect of conversational style if

- either the language features forms that *directly* correspond to its realization (★),
- or it is possible to use a pattern or encoding on a *local* basis [25, 41] that provides such support and is faithful to the model (☆).

For example, Active Object languages support concurrency *within* an active object meaning it allows a single active object to conduct simultaneous conversations (=).

By contrast, most concurrency models may achieve a form of support for branching subconversations ($\boldsymbol{\propto}$) by delegating to spawned threads/actors/processes. Doing so requires additional coordination between the parent and child components, but this coordination can often be encapsulated through the use of a suitable pattern or abstraction. Moreover, the use of such an abstraction is *local* in the sense that only the part of the program where it is employed requires any knowledge of its existence.

A blank cell in the table indicates that language implementations of a model fail to satisfy either criteria. See Appendix B for additional discussion of the table cells.

# 7 Conclusion

The communication among concurrent components in a program forms a conversation. Programs with numerous such components naturally feature a number of such conversations. As a program evolves, these conversations progress in both parallel and sequential fashions. Moreover, conversations may give rise to related subconversations, each of which exhibits the same possibilities for evolution. The implementation of individual components must express these shifting possibilities as directly as possible.

This paper argues Syndicate's pragmatic capabilities for expressing conversational concurrency with an extended example. Syndicate achieves this goal by combining dataspaces for maintaining conversational context with facets for organizing an actor's internals. The extended example provides a basis for comparing Syndicate with other approaches with respect to conversational concurrency, and the comparison indicates that Syndicate is uniquely capable in this regard.

**Acknowledgements** We would like to thank the anonymous reviewers for their feedback and advice. This work was partially supported by NSF grant SHF 2315884.

# A Racket, the Language

Racket is a descendant of the Scheme programming language and, as such, a member of the Lisp family. While all these languages come with a standard procedural, lexically-scoped core programming language, Racket also serves as a research vehicle for rapidly prototyping languages such as Syndicate. As far as this paper is concerned, Racket's core comes with the usual contemporary linguistic constructs of a mostly-functional language: structures and first-class instances; first-class and higher-order functions; pattern matching; but also mutable data structures and assignable variables. For prototyping languages, Racket features an expressive macro system for seamlessly extending the language with new constructs [21] and even domain-specific languages [26].

**Core Racket** Like all members of the Lisp family, Racket sports a fully parenthesized, prefix-operator syntax. As a consequence, the grammar of variables and simple strings (called symbols) accommodates many more keyboard characters than in standard syntaxes; it excludes only white space characters. For example, `stop-with!` is a perfectly legal identifier; Racket programmers use `!` for a function name if it realizes an effect (assignment, jump). Similarly, a single quote character at the beginning of such a character sequence denotes a simple string; `'no-price-match` is an example.

All plain expressions start with `(` followed by an operator:

```
(>= price-offered desired-price)

(- held-funds (* desired-no-shares actual-price))
```

The first expression checks whether the value of `price-offered` is larger or equal to the value of `desired-price`, while the second expression subtracts the product of `desired-no-shares` and `actual-price` from `held-funds`. Due to this highly uniform syntax, Racket's primitive operators are merely pre-defined functions.

Programmers can define functions, using the following Racket form:

```
(define (function-name parameter-1 ... parameter-N)
  definitions-and-expressions-1
  ...
  definitions-and-expressions-N
  expression)
```

The scope of such definitions is the entire module or, if they are located within the definition's body, it is just the function body. In addition to named functions, Racket also supports higher-order, nameless functions using Church's original notation:

```
(lambda (parameter-1 ... parameter-n) expression).
```

Unsurprisingly, applying such defined functions uses the same syntax as the use of primitive operators:

```
(function-name argument-1 ... argument-N)
```

To make all this concrete, here is a function that compares an actual price of an item with an offered price, which is defined in the context of the function definition:

```
(define price-offered ...)

#; { Real -> Boolean }
(define (acceptable-price actual-price)
  (>= price-offered actual-price))
```

The commented line (`#;`) is suggestive of a type-like signature. By contrast, the following function definition uses an inner definition to extract the price of an offer from a given order:

```
(define (work-on-one-order the-order actual-price)
  (define price-offered (select-price the-order))
  (>= price-offered actual-price))
```

The scope of the bold-faced definition is just the two lines of the function body (lines 2 and 3).

The above function definition assumes more than the usual basic data—numbers, Booleans (`#true`, `#false`), etc. Racket programmers can introduce structured data via struct-type definitions:

```
1    (struct name-of-struct (name-of-field-1 ...name-of-field-N))
```

Such a definition generates `N` plus two functions: a constructor named `name-of-struct`, a predicate ("instance of" check) named `name-of-struct?`, and `N` selectors: `name-of-struct-field-1` through `name-of-struct-field-N`. Let's make this concrete with two struct definitions:

```
1    (struct trading-day-open ())
2
3    (struct order-result (order result))
```

The first one comes with a constructor that takes no arguments, meaning an instance of this structure type is generated with (`trading-day-open`). In contrast, the second struct-type definition injects four functions into a programmer's code: `order-result`, `order-result?`, `order-result-order`, and `order-result-result`. Thus,

```
1    (define or1 (order-result x 'no-purchase))
2
3    (if (> (order-result-order or1) 2)
4        (order-result-result or1)
5        'ouch)
```

creates an instance of the struct type (from some value `x` and `'no-purchase`) and then checks whether its `order` field is larger than 2. If so, it extracts the `result` field from `or1`; otherwise the result of the condition expression is `'ouch`.

An alternative way to extract values from a struct uses algebraic pattern matching:

```
1    #; { (InstanceOf OrderResult) -> ... }
2    (define (deal-with-result ordr)
3      (match-define (order-result o r) ordr)
4      ... o ... r ...)
```

As indicated by the type-like comment, this function consumes an instance of the `order-result` struct type. The pattern-matching define makes the two field values available in the remaining definitions and expressions of the function as `o` and `r`.

*Note on Convention.* In this presentation, we sometimes use the name of a struct's constructor to refer to an instance, or instances of, the struct. For example, we may say "an `order-result` indicates that all work on the order is complete" to detail a program's usage of values that are instances of the struct rather than the constructor itself. The context of the statement should clarify the intended meaning.

**Programmable Racket** The Racket-way of experimenting with language design is to re-program Racket [26]. The result of such a re-programming effort is a full-fledged programming language: (1) as performant as Racket; (2) as rich in basic data and functionality as Racket; (3) and as equipped with libraries, including a graphical UI framework, as the base language. Indeed, even the standard IDE, DrRacket [27], adapts itself to such a language.

Re-programming Racket has two distinct meanings. The first one may include changing the semantics of some or all of Racket's constructs. For example, Felleisen et al. [26] sketch how to assign a lazy semantics to the conventional syntax of Racket.

The interested reader may also check out the work of Tobin-Hochstadt et al. [52], which explains how to equip Racket with types.

The second meaning is Lisp's conventional idea of adding syntactic constructs to the language's core grammar. The result is often called a *hosted language*. Such a hosted language is a seamless extension of the host, both in terms of syntax and semantics. Working with a hosted language, a programmer will soon forget which syntactic forms come from core Racket and which ones from, say, Syndicate. Programming in this new language quickly feels as natural as programming in plain Racket.

The difference between Racket and other Lisp languages is that syntactic constructs are defined in modules, exported from modules, and imported into modules. Racket also permits the definition of context-dependent syntax forms—such as the above-mentioned type system for Racket—and even cross-module forms [29].

A syntactic language extension is defined via a set of macros. Roughly speaking, a macro is a rule for rewriting syntactic patterns into expressions. Here is an example:

```
1  (define-syntax-rule
2    #; { THE SYNTACTIC PATTERN }
3    (when-not expression
4      defn-or-expr-1 ... final-expr)
5    #; { TRANSLATES INTO THIS CORE RACKET EXPRESSION: }
6    (if expression
7        #; { THEN }
8           (eprintf "warning: when-not failed~n")
9        #; { ELSE }
10          (let ()
11            defn-or-expr-1 ... final-expr)))
```

It defines a new syntactic form, dubbed `when-not`, which evaluates a series of definitions and expressions if the first expression evaluates to `#false`. The definition takes the form of a rewrite rule, stating a *pattern* (lines 3 and 4) that translates into a *template* (lines 5 through 11). The `defn-or-expr-1 ...` pattern combines a pattern variable (`defn-or-expr-1`) with the ... convention [55] to indicate that a potentially empty sequence is grammatically correct in this position. The final pattern variable (`final-expr`) indicates that the last S-expression in the body of `when-not` must be a plain expression. If any S-expression matches the pattern during compilation, the Racket compiler replaces it with the template, after substituting the pattern variables with their actual matched pieces [55]. If such a match concerns a sequence of expressions, the template is filled with the entire sequence. While Racket was originally based on Wand and Kohlbecker's macro-by-example system, it comes with several powerful extensions for writing such definitions [6, 20, 21, 23]; additionally, Racket also uses hygienic macro expansion [17, 18, 28, 38] to avoid interference between identifiers from the use site of the macro with those in the definition site.

Creating an extension of Racket like the one for Syndicate requires programming a module that exports the additional syntactic constructs and functions. A programmer may then specify this new language at the top of a module with a `#lang` declaration:

```
1    #lang syndicate
```

The module's code will freely mix Racket's core syntax with Syndicate's extensions.

## B Extended Discussion

This Appendix expands on the assignment of ✩, ★, and empty cells from Figure 10.

Actors, Active Objects, Channels, and Sparrow share the same rationale for partially supporting conversations with evolving groups of participants (👥) and flexible notions of identity (∞). In each language, a developer may implement a single coordinating actor representing the group itself. An actor messages the coordinator to register itself to join a conversation and then again to leave. To send a message to the current group, an actor messages the coordinator, which duly relays it to the current participants. A similar strategy allows looking up the identity of an actor (∞). A correct implementation of the coordinator handling the many possible error cases is no mean task. However, once done, it seamlessly integrates with the existing idiom.

Likewise, most languages receive partial credit for supporting branching conversations (branch) due to the ability to delegate a subconversation to a new actor, process, or thread. The parent and child then communicate as necessary to keep each conversation in sync. This synchronization logic may grow increasingly complex depending on the relationship between the conversations. In Fact Spaces, concurrency occurs at a per-device level in the system, limiting the ability to employ such a strategy.

Tuples directly represent the conversational context (🗄) of the program. In Lime, programmers specify reactions that activate some behavior in response to the appearance of a tuple. Fact Spaces and AmbientTalk provide similar capabilities. Note that the connection between context and behavior is only one way, from tuple to reaction. By contrast, dataspaces and facets provide a bidirectional connection. The `during` form activates and deactivates a facet based on the presence of an assertion. In the other direction, a facet's assertions are automatically incorporated into the conversational context when it is active and removed when it is gone.

Communication in tuplespace languages, including Fact Spaces and AmbientTalk, is also agnostic to the participants in a conversation (👥) and their identities (∞). Recipients receive messages based on patterns over tuples, and may choose not to consume the tuple, allowing other components to receive it as well.

**Channels** A process can weave together simultaneous conversations (⇌) by performing a `sync` or `select` operation across all channels on which a message may appear. The process handles the first event that is ready and repeats. The logic to maintain control over many active channels at a single point of control may grow complex, however.

Channels reify elements of conversational context (🗄). Specifically, each individual channel represents a (sub)conversation within the program. However, there is little support for managing the context or connecting it with the appropriate behaviors.

**Multithreading with Shared Memory** Certain patterns of conversational parallelism (⇌), such as fork/join communication patterns, have dedicated linguistic support, and the memory shared between threads is the program's context (🗄). Threads may easily access and update the context. One of the primary challenges of this style of concurrency is employing judicious synchronization to keep the context consistent.

Condition variables indirectly link thread behavior to certain conversational contexts, but this connection must be managed manually.

**Sparrow** Sparrow improves upon the conversational support provided by Erlang actors. Reactions allow managing the behavior of each conversation separately (⚌). It lacks a structuring mechanism beyond a single pattern or reaction. Similarly, the messages received over time by an individual actor represent a particular slice of the context (🗄). Complex event patterns directly express the conditions on this context that a particular behavior requires to activate, and then again to deactivate.

## C  State Machine Notational Definition

Figure 11 explains how `state-machine` is a shorthand for a facet that gets re-started—via a `stop` and its continuation behvaior—whenever the machine transitions from one state to another.

```
(state-machine name
  (label-0
   body-0 ...)
  ((label-N argN ...)
   body-n ...)
  ...)
```

$\stackrel{\text{def}}{\equiv}$

```
(define (run-state state args)
  (state-machine
    (set! id (current-facet-id))
    (match state
      (mark-0 body0 ...)
      (mark-N
       (match-define (list argN ...) args)
       bodyN ...)
      ...)))

(define id #false)

(define (goto name label-K . args')
  (stop id (run-state mark-K args')))

(run-state label-0 (list))
```

where `mark-0` through `mark-N` are unique values corresponding to `label-0` through `label-N`; and where `id` is an identifier not used in the rest of the program.

■ **Figure 11** The `state-machine` abstraction

## D  Additional Implementation Examples

The implementation of the market scenario in Section 4 is incomplete. It needs some additional, synthetic actors, meaning actors that do not have a direct correspondence in the scenario.

■ **Listing 6** Implementation of the wallet actor

```
1  (define (spawn-wallet)
2    (spawn
3      (field (processed (set)))
4      (on (asserted (funds-needed $the-order $account $amt))
5          (unless (set-member? (processed) the-order)
6            (processed (set-add (processed) the-order))
7            (react
8              (define parent (current-facet-id))
9              (assert (withdraw-funds the-order account amt))
10             (during (bank-response the-order $ok?)
11               (assert (funds-held the-order account amt ok?))
12               (on (asserted (order-result order $ans))
13                   (stop parent
14                         (when (and ok? (not (equal? 'fulfilled ans)))
15                           (deposit! account amt)))))))))
16
17 (define (deposit! account amt)
18   (unless (zero? amt)
19     (spawn
20       (define deposit-txn-id (gensym))
21       (assert (deposit-funds deposit-txn-id account amt))
22       (stop-when (asserted (bank-response deposit-txn-id _))))))
```

### D.1 Wallet

Section 4.5 mentions the wallet actor, which the broker actor spawns to act as its transparent intermediary with the bank. Using an intermediary addresses two potential problems:

1. Maintaining the context of an order across trading days, especially the funds held on behalf of a buyer.
2. Handling the scenario where the buyer cancels its order while there is an outstanding request to withdraw funds from the bank.

Listing 6 presents the complete definition of wallet. The `processed` field (line 3) keeps track of which orders the wallet actor has seen before. When the broker first needs funds for an order, the wallet actor creates a facet (line 7) to handle the related banking interaction. The facet asserts a request to withdraw funds (line 9) and then waits for the response (line 10). Critically, the facet does not respond to any events until a response from the bank arrives. Once available, the facet provides a `funds-held` assertion (line 11) for the broker based on the response from the bank. Finally, when the broker completes the order (line 12), the facet shuts down and returns any leftover funds (lines 13–16).

## E  Syndicate/JS

The JavaScript/TypeScript implementation ("Syndicate/JS") exists as a transpiler that compiles Syndicate-specific syntax extensions to unadorned TypeScript/JavaScript. The implementation augments the underlying language with direct support for Syndicate's dataspace patterns, as well as with support for Syndicate's `spawn`, `during`, `on asserted` statements, and so on.

The purpose of this section is to give readers an impression of what Syndicate code looks like in this implementation. As in the preceding sections, syntactic Syndicate extensions are underlined. Listing 7 shows the syntax for declaring the "market" protocol in Syndicate/JS. These structure type declarations translate to plain "record" definitions in the underlying language.

**Listing 7**  Structures used in the protocol for the simple market, in Syndicate/JS.

```
assertion type TradingOpen();
assertion type Order(id, account, amt, price);
assertion type OrderResult(order, answer);
assertion type FundsNeeded(id, account, amt);
assertion type FundsHeld(id, account, amt, isOk);
assertion type Price(v);
assertion type PurchaseRequest(id, amt, p);
assertion type PurchaseResponse(id, isOk);
```

Listings 8 and 9 illustrate how programmers can express the "seller" and "broker" actors, respectively, in Syndicate/JS. The code in Listing 8 is a direct translation of Figure 6.

**Listing 8**  Syndicate/JS definition of the seller.

```
function spawnSeller(sharePrice) {
 spawn {
  during TradingOpen() => {
   assert Price(sharePrice);
   during PurchaseRequest($id, $amt, $p) => {
    assert PurchaseResponse(id, p >= sharePrice);
}}}}
```

By contrast, Listing 9 integrates several code snippets from the main body of the paper into a single broker function. Specifically, the Syndicate/JS code inlines the code from Figures 7, 8, and 9 into a `spawnBroker` function.

▪ **Listing 9** Syndicate/JS definition of the broker.

```
function spawnBroker() {
 spawn {
  during TradingOpen() => {
   on asserted Order($id, $account, $numShares, $desiredPrice) =>
   react {
     const theOrder = Order(id, account, numShares, desiredPrice)
     const stopWith = makeStopWith(theOrder, currentSyndicateFacet);
     const potentialCost = numShares * desiredPrice;
     react {
      assert FundsNeeded(theOrder, account, potentialCost);
      on retracted theOrder => stopWith(CANCELED);
      on asserted FundsHeld(theOrder, account, potentialCost, false) => {
       stopWith(INSUFFICIENT_FUNDS);
      }
      stop on asserted FundsHeld(theOrder, account, potentialCost, true) =>
       react {
        on asserted Price($actualPrice) => {
         if (actualPrice > desiredPrice) {
           stopWith(NO_PRICE_MATCH);
         } else {
           assert PurchaseRequest(theOrder, numShares, actualPrice);
           on asserted PurchaseResponse(theOrder, true) => {
            const leftover = potentialCost - (numShares * actualPrice);
            returnFundsAndStop(account, leftover, FULFILLED, stopWith);
         }}}
         on retracted theOrder => {
          returnFundsAndStop(account, potentialCost, CANCELED, stopWith);
}}}}}}}
```

## About the authors

**Sam Caldwell** Contact Sam at samc@ccs.neu.edu.
https://orcid.org/0000-0001-7092-8769

**Tony Garnock-Jones** Contact Tony at tony.garnock-jones@maastrichtuniversity.nl.
https://orcid.org/0009-0000-9516-0848

**Matthias Felleisen** Contact Matthias at matthias@ccs.neu.edu.
https://orcid.org/0000-0001-6678-1004